\documentclass[aps,reprint,amsmath,amssymb, notitlepage,onecolumn,longbibliography,showpacs,pra]{revtex4-1}
\usepackage{graphicx}% Include figure files
\usepackage{bm}% bold math
\usepackage{hyperref}

\begin{document}
	
	\title{Spin hydrodynamic generation in unsteady flows}
	\author{Takumi Funato${}^{1,2}$}
	\author{Mamoru Matsuo${}^{2,3,4,5}$}
	\email{gmtsuo@gmail.com}
	\affiliation{$^1$Center for Spintronics Research Network, Keio University, Yokohama 223-8522, Japan}
	\affiliation{$^2$
		Kavli Institute for Theoretical Sciences, University of Chinese Academy of Sciences, Beijing, 100190, China.
	}
	\affiliation{$^3$
		CAS Center for Excellence in Topological Quantum Computation, University of Chinese Academy of Sciences, Beijing 100190, China
	}
	\affiliation{$^4$
		Advanced Science Research Center, Japan Atomic Energy Agency, Tokai, 319-1195, Japan
	}
	\affiliation{$^5$RIKEN Center for Emergent Matter Science (CEMS), Wako, Saitama 351-0198, Japan}

	\date{\today}
	
	\begin{abstract}
	    We theoretically investigate a spin-mediated conversion from fluid dynamics to voltage, known as spin hydrodynamic generation (SHDG), in oscillatory and transient unsteady flows.
	    We consider unsteady flows of liquid metal between two parallel infinite planes and then calculate its vorticity fields based on the Navier--Stokes equation for an incompressible viscous fluid.
	    The spin accumulation and spin current generated by unsteady flows are derived using a spin-diffusion equation, including spin-vorticity coupling, which is a couple of angular momentum between electron spin and vorticity field in unsteady flows.
	    The estimation of SHDG in liquid mercury flow suggests that an observable magnitude of voltage can be induced in unsteady flows.
	    Our results are expected to enable the realization of high-speed spin devices with unsteady flows and broaden the range of fluid spintronics applicability.   
	\end{abstract}

	\pacs{Valid PACS appear here}
	\maketitle
	
	\section{Introduction}
Spin hydrodynamic generation (SHDG)~\cite{takahashi2016Spin,matsuo2017Theory,takahashi2020Giant,tabaeikazerooni2020Electron,tabaeikazerooni2021Electrical,tokoro2022Spin} is an electric voltage generation mediated by a spin current~\cite{matsuo2011Effects}, a flow of electron spin angular momentum, driven by the coupling between fluid vorticity and electron spins. 
SHDG interconnects the apparently unrelated two research fields, fluid mechanics and spintronics, in a close relationship.

	SHDG's origin is spin-vorticity coupling (SVC)~\cite{matsuo2011Effects,matsuo2011Spin,matsuo2017SpinMechatronics,matsuo2017Theory}, an interconversion between mechanical angular momentum associated with the vorticity of fluid and electron spin angular momentum.
	In the SHDG mechanism, an angular momentum associated with the vorticity of fluid motion generates a spin current via SVC. 
	Thus, this induced spin current is converted to voltage via the inverse spin Hall effect~\cite{saitoh2006Conversion,valenzuela2006Direct}, which is a spin-to-charge conversion via the spin--orbit interaction.
	The first SHDG observation was reported by Takahashi {\it et al}~\cite{takahashi2016Spin}. 
	They experimentally confirmed the generation of charge current in the streamwise direction of turbulent flows of liquid metal, specifically mercury and galinstan (GdInSn), through quartz capillaries.
	The velocity dependence of the inverse spin Hall voltage in laminar and turbulent flows was theoretically demonstrated based on the quantum kinetic theory~\cite{matsuo2017Theory}. 
	Thereafter, the predicted velocity dependence of SHDG has been experimentally confirmed for laminar and turbulent flows of mercury in the Reynolds number range of $10<\text{Re}<10000$~\cite{takahashi2020Giant}. 
	In addition, the SHDG efficiency for the laminar flow is much greater than the turbulent flow.
	Kazerooni {\it et al.} experimentally confirmed SHDG in pipe flows of liquid galinstan at low and high Reynolds numbers and obtained conclusions consistent with a previous report~\cite{tabaeikazerooni2020Electron}.
	The system configuration dependence of SHDG was investigated experimentally for the laminar flow in circular pipes and rectangular and square ducts.
	The results indicate that their SHDG efficiency is the same when the diameter and height are the same~\cite{tabaeikazerooni2021Electrical}.
	The SHDG in liquid gallium, whose spin--orbit interaction is weaker than mercury, has recently been observed~\cite{tokoro2022Spin}.
	As mentioned previously, the physical properties of SHDG in steady flows, such as laminar and turbulent flows, have been investigated intensively using experimental and theoretical approaches.

		\begin{figure}
	    \centering
	    \includegraphics[width=100mm]{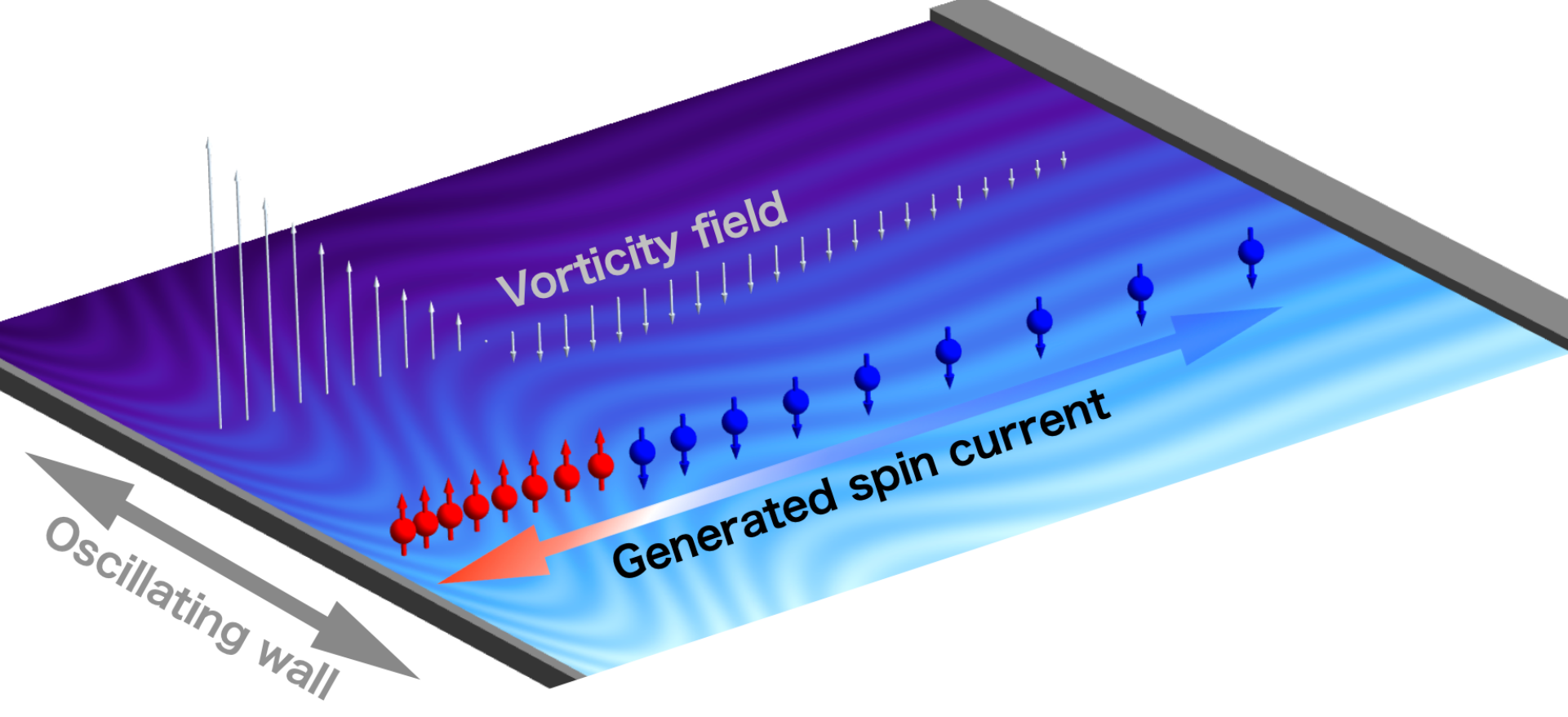}
	    \caption{
	    (Color online)
	    Schematic illustration of spin-current generation in unsteady flows via SVC.}
	    \label{fig:1}
	\end{figure}

Furthermore, SVC is utilized in the angular momentum conversion between electron spin and vorticity of a surface acoustic wave (SAW) (i.e., an acoustic wave transmitted along the plane surface of a material). 
In this mechanism, local lattice rotation associated with a SAW drives spin currents via SVC~\cite{matsuo2013Mechanical,matsuo2017SpinMechatronics,kobayashi2017Spin,kurimune2020Highly,kurimune2020Observation,tateno2020Electrical,tateno2020Highly,funato2018Generation,kawada2021Acoustic,funato2021Acoustic,funato2021Helicity,funato2022Spina}. 
Along with research on conventional SAW-driven magnetization physics~\cite{uchida2008Observationa,uchida2011Longrange,uchida2011Surfaceacousticwavedriven,weiler2011Elastically,weiler2012Spin,keshtgar2014Acoustic,deymier2014Phononmagnon,adachi2014Theorya},  
 these studies form a major field of SAW spintronics.
	SAW-driven spin current is implemented in a high-frequency region on the order of GHz, near magnetic resonance frequency. 
	Thus, SAWs might be used to realize high-speed spintronics devices, while only fluid spintronics has been investigated in the steady state regime of the flow, namely, the time regime longer than the decay time of fluid dynamics.
	The spin-mediated phenomena should be clarified due to unsteady flow to widen the range of fluid spintronics applicability; however, this remains an open problem.
	
	This study aims to theoretically investigate SHDG in two types of unsteady flows (i.e., oscillatory and transient) of liquid metal (Fig.~\ref{fig:1}).
	We consider liquid metal enclosed between two infinite parallel plates and generate an oscillatory flow because of the in-plane harmonic motion of one plane and a transient flow because of the sudden constant motion of one plane. 
	The velocity and vorticity fields of the two flow types are determined by using the Navier--Stokes equation of an incompressible viscous fluid. 
	Meanwhile, the spin accumulation and spin current induced by the unsteady flows via SVC are calculated by using the spin-diffusion equation.
	Considering liquid mercury, we estimate the inverse spin Hall voltage converted from the spin current generated by unsteady flows.
	We found that unsteady flows induce an observable magnitude of voltage.
	Our results are expected to enable the realization of high-speed fluid spin devices with unsteady flows and expand the fluid spintronics region.

	\section{spin-diffusion equation}
	This section introduces the spin-diffusion equation that describes the spin transport in a liquid metal flow. The spin transport in liquid metal is mainly contributed by conduction electrons, the same as that in solid metal, and here we focus on conduction electrons. 
	The motion of conduction electrons subjected to the ionic potential in liquid metal at rest is governed by the following Schr\"odinger equation:
	\begin{align}
	    i\hbar \frac{\partial }{\partial t} \psi (\bm x,t) = \left( -\frac{\hbar^2 }{2m} \frac{\partial ^2}{\partial x^2} + U(\bm x) \right)\psi (\bm x,t),
	\end{align}
	where $\psi(\bm x,t)$ is the spinor of the conduction electrons, and $U(\bm x)$ is the potential due to the ions.

	When liquid metal flow with the velocity profile $\bm u(\bm x,t)$ is induced, the inertial effect that acts on the electron spin arises, which is the coupling between a mechanical rotation and an electron spin.
	We perform the coordinate transformation from the laboratory frame $\bm x$ to an inertial frame fixed to the flowing ions, $\bm r=\bm x - \bm vt$, to treat the inertial effect because of the viscous fluid flow for simplicity.
	The electrons are assumed to adiabatically follow the flow of the viscous fluid, that is, the mean-free path $l$ and relaxation time $\tau$ of the conduction electrons are much shorter than the spatial and temporal variations of the velocity profile, respectively. They are represented as $l |\nabla u/u|\ll 1$ and $\tau |\partial_t u/ u|\ll 1$.
	The Schr\"odinger equation that governs the motion of the conduction electrons in the flowing fluid is given by
	\begin{align}
	    i\hbar \frac{\partial}{\partial t} \psi(\bm r,t) = \left( -\frac{\hbar^2}{2m}\frac{\partial^2}{\partial r^2} + U(\bm r) - \frac{1}{2} \bm S \cdot \bm \Omega (\bm r,t) \right) \psi(\bm r,t),
	    \label{eq:svc}
	\end{align}
	where $\bm S=\hbar \bm \sigma /2$ is the spin angular momentum operator with the Pauli matrices $\bm \sigma =(\sigma_x,\sigma_y, \sigma_z)$ and $\bm \Omega = \nabla \times \bm v$ is the vorticity.
	The third term $-\bm S \cdot \bm \Omega/2$ is called SVC, which reproduces the spin-rotation coupling $-\bm S \cdot \bm \Xi$ with the rotation frequency $\bm \xi$ in analogy with the vorticity $\bm \Omega$ reduced to the rotation frequency $\bm \Xi$ as $\bm \Omega =2\bm \Xi$ for a rigid-body rotation.
	
	According to the quantum kinetic theory, the spin transport in the electron system satisfying Eq.~(\ref{eq:svc}) is governed by the following spin--diffusion equation:
	\begin{align}
	    \left(
	        \frac{\partial}{\partial t} -D \frac{\partial^2}{\partial r^2} + \tau_{\text{sf}}^{-1}
    	 \right) \delta \mu_s
	    = - \frac{\hbar \dot{ \Omega}}{4} - \hbar \tau_{\text{sf}}^{-1} \zeta { \Omega},
	    \label{eq:spin_diff}
	\end{align}
	where $\delta \mu_s=(\mu_{\uparrow}-\mu_{\downarrow})/2$ is the spin accumulation with $\mu_{\uparrow}$($\mu_{\downarrow}$) being the chemical potentials of up-spin (or down-spin) conduction electrons, $D$ is the diffusion constant, $\tau_{\text{sf}}$ is the spin relaxation time, and $\zeta$ is the renormalization factor.
	%, which was estimated as $\zeta \sim 10^2$ in liquid metal flow
	Here, the direction of spin polarization is chosen along the vorticity direction.
	The right-side terms in Eq.~(\ref{eq:spin_diff}) correspond to the spin-source terms because of the SVC.
	The spin current is driven by the diffusion of spin accumulation induced via SVC and is described by the gradient of spin accumulation:
	\begin{align}
	    \bm j_{\text s}(\bm r) = \frac{\sigma_{\text c}}{e} \frac{\partial \delta \mu_s}{\partial \bm r},
	\end{align}
	where $\sigma_{\text c}$ is the conductivity and $e(>0)$ is the charge elementary.
	Note that the spin current is defined to have the same dimension as the charge current density A/m$^2$.
	We emphasize that the gradient distribution of the vorticity requires the spin current generation through SVC.

    \section{Spin current driven by unsteady flow}
    
    We examine the spin current generation by an unsteady flow of liquid metal.
    The motion of an incompressible viscous fluid is governed by the Navier--Stokes equation:~\cite{book1}
    \begin{align}
	    \frac{\partial \bm u}{\partial t} + \left(\bm u\cdot \frac{\partial}{\partial \bm r} \right) \bm u = -\frac{1}{\rho} \frac{\partial p}{\partial \bm r} + \nu \frac{\partial^2 \bm u}{\partial r^2},
	\end{align}
	where $\bm u$ is the velocity of the fluid, $p$ is the pressure, and $\nu=\eta/\rho$ is the kinetic viscosity, which is the ratio of the viscosity coefficient $\eta$ to the fluid density $\rho$.
	
	\begin{figure}
	    \centering
	    \includegraphics[width=120mm]{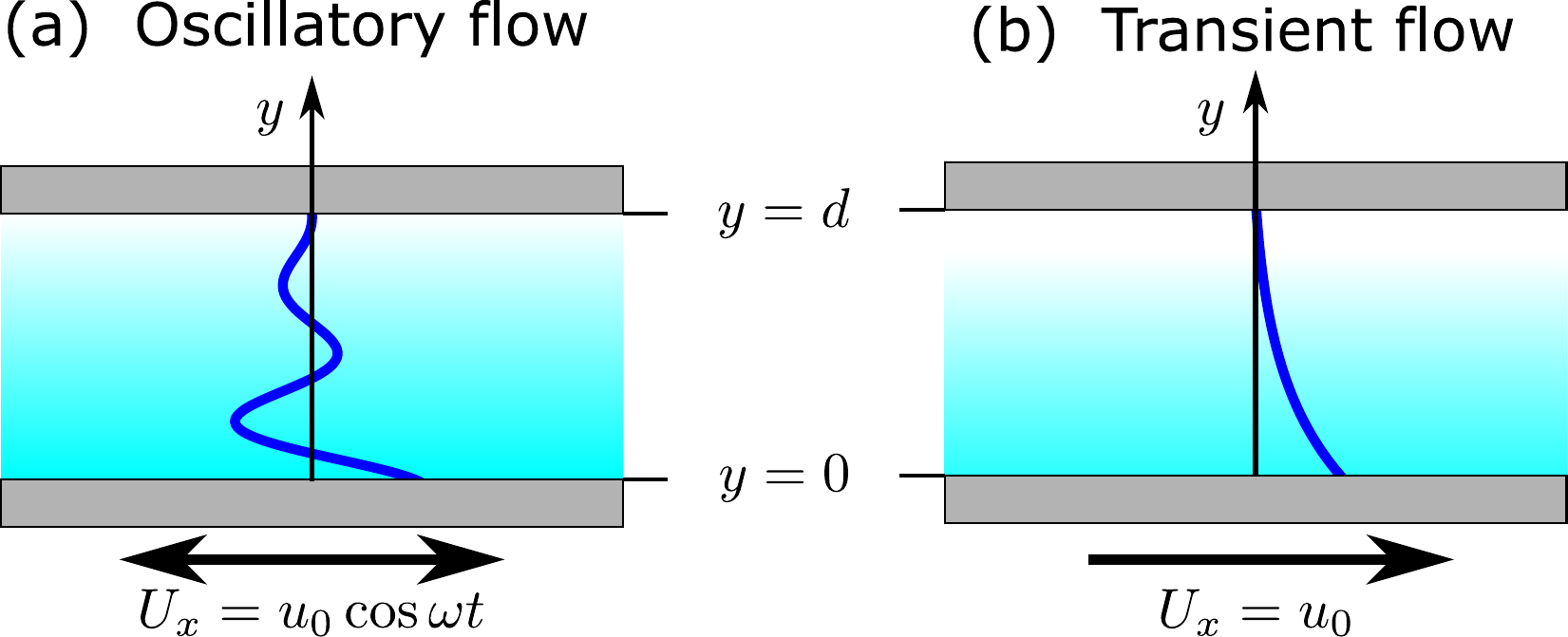}
	    \caption{
	    (Color online)
	    Schematic illustration of the oscillatory and transient flow of liquid metal.
	    (a) The oscillatory flow generated by the harmonic oscillation of the bottom plate and (b) the transient flow generated by the sudden translation motion of the bottom plate.
	    The blue parts represent the velocity profiles of liquid metal.
	    }
	    \label{fig:setup}
	\end{figure}
	
	\subsection{Oscillatory flow}
\begin{figure}
	    \centering
	    \includegraphics[width=160mm]{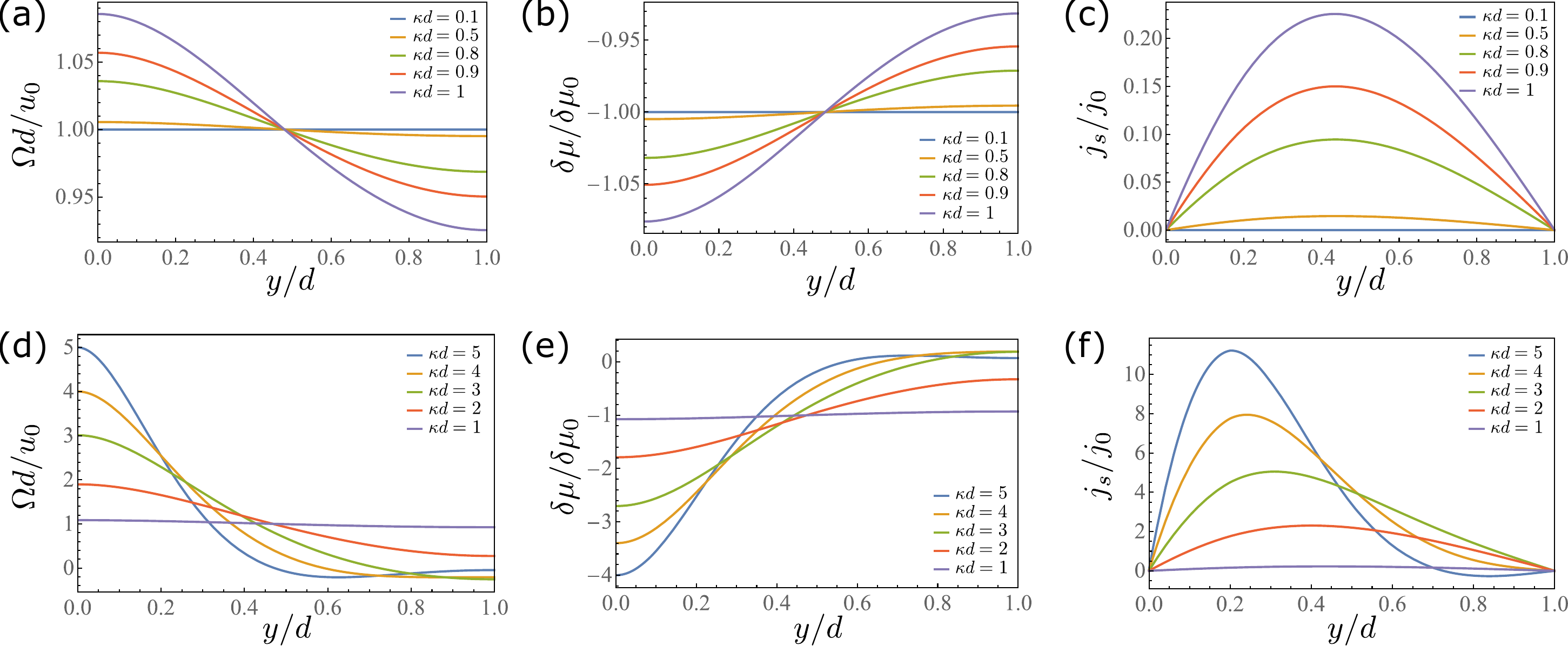}
	    \caption{
	    (Color online)
	    Spatial distribution of vorticity, spin accumulation, and spin current generated by the oscillatory flow.
	    (a-b) The normalized amplitude of the oscillating vorticity, (c-d) the amplitude of the spin accumulation normalized by $\delta \mu_0=\hbar \zeta u_0 /d$, and (e-f) the amplitude of the spin current normalized by $j_0 = \sigma_c \hbar \zeta u_0/ed^2$.
	    Colored lines represent the ratio of the boundary layer thickness to the distance between the two plates.
	    The vorticity and spin accumulation become spatially uniform as $\kappa d$ becomes smaller and the generated spin current becomes small.
	    }
	    \label{fig:osc}
	\end{figure}

    Let us consider incompressible viscous fluids enclosed between two infinite parallel plates with distance $d$.
    The upper plane at $y=d$ is fixed, while the bottom plane at $y=0$ oscillates harmonically in the $x$-direction having a velocity $U_x(t) =u_0\cos \omega t$.
    In this case, the oscillatory Couette flow is produced by the oscillating plane.
	As the unidirectional flow of the viscous fluid is driven in the present setup, its velocity profile is parallel to the $x$-direction and independent of $x$ and $z$ (i.e., $\bm u(\bm r, t) = u(y,t)\hat x$), with $\hat x$ being the unit vector along the $x$-direction.
	The Navier-Stokes equation can be solved analytically with the boundary conditions $u(0,t)=u_0\cos \omega t$ and $u(d,t)=0$ as
	\begin{align}
	    u(y,t) = \text{Re}\, \left[ u_0 \frac{\sinh \xi (d-y)}{\sinh \xi d} e^{i\omega t} \right],
	\end{align}
	where $\xi=\kappa (1+i)$  with $\kappa =\sqrt{\omega/2\nu}$ being the relaxation constant of the oscillatory flow.
	The fluid accompanies the oscillating plate in the thickness of the boundary layer $\kappa^{-1}$.
	The vorticity associated with the oscillatory flow occurs parallel to the $z$-direction and varies along the $y$-direction  (i.e., $\bm \Omega (\bm r,t)= \Omega (y,t)\hat z =-\partial _y u(y,t) \hat z$) and is given by
	\begin{align}
	    \Omega (y,t) = \text{Re}\, \left[ u_0 \frac{\xi \cosh \xi (d-y)}{\sinh \xi d} e^{i \omega t} \right].
	    \label{eq:omega_oscillatory}
	\end{align}
	Figure.~\ref{fig:osc} (a) and (d) represent the plot of the normalized amplitude of the oscillating vorticity.
	The spatial vorticity variation decreases evidently as the boundary thickness becomes longer than the distance between the two plates.

	We substitute the vorticity given in Eq.~(\ref{eq:omega_oscillatory}) to the spin-diffusion equation given in Eq.~(\ref{eq:spin_diff}) to calculate the spin current driven by the oscillatory flow of liquid metal.
	When the oscillatory motion of the bottom plane is much slower than the characteristic time of spin relaxation (i.e., $\omega \tau_{\text{sf}}\ll 1$), the time derivative term in the right-hand side of Eq.~(\ref{eq:spin_diff}) becomes negligible, and the spin-diffusion equation becomes approximately steady.
    Furthermore, the first term of the spin-source term on the right-hand side of Eq.~(\ref{eq:spin_diff}), which is contained by the time derivative of vorticity, becomes negligible when compared with the second term of the spin-source term. Therefore, the spin accumulation induced by the oscillatory flow is determined by the following spin-diffusion equation:
	\begin{align}
	    \left( \lambda^2\frac{\partial^2}{\partial y^2} -1 \right) \delta \mu_s(y,t) = \hbar \zeta \Omega(y,t),
	    \label{eq:spin_diff_osc}
	\end{align}
	where $\lambda =\sqrt{D\tau_{\text{sf}}}$ is the spin-diffusion length.
    	With the boundary conditions that the spin current at the top and bottom plates vanish (i.e., the $y$-directional derivative of the spin accumulation at the plates vanish $\partial_y \delta \mu_s(y=0)=\partial_y \delta \mu_s(y=d)=0$), we solve the steady spin-diffusion equation, Eq.~(\ref{eq:spin_diff_osc}).
	Figure~\ref{fig:osc} (c) and (d) indicate the spatial distribution of the amplitude of the oscillating spin accumulation normalized by $\delta \mu_0=\hbar \zeta u_0 /d$.
	Figure~\ref{fig:osc} (e) and (f) represent the spatial distribution of the amplitude of the spin current normalized by $j_0 = \sigma_c \hbar \zeta u_0/ed^2$, where we assume that $\lambda /d=0.1$.
	The spin accumulation becomes spatially uniform as the boundary layer thickness becomes longer than the distance between the two plates (i.e., $\kappa d$) and the generated spin current becomes small as shown in Fig.~\ref{fig:osc}.
	
	These boundary layer thickness dependencies can be suggested by the intuitive picture of the vorticity associated with the oscillatory flow for the two limits: the boundary thickness $\kappa^{-1}$ is much shorter than the interplate distance $d$ (fast oscillation) and  $\kappa^{-1}$ is much shorter than $d$ (slow oscillation).
	The vorticity in the two cases is approximately given by
	\begin{align}
	    \Omega(y,t) \simeq 
	    \begin{cases}
	        (u_0 \kappa e^{-\kappa y}/\sqrt 2) \cos (\omega t-\kappa y+\pi/4) & (1 \ll \kappa d),\\
	        (u_0 /d) \cos \omega t & (1 \gg \kappa d).
	    \end{cases}
	\end{align}
	In the first case, the gradient distributed vorticity occurs along the decay of the oscillatory flow parallel to the $y$-axis. 
	The spin accumulation has a gradient distribution via the SVC, and the diffusion spin current can be driven. 
	In the second case, the vorticity has a spatially uniform distribution.
	Therefore, an oscillatory Couette flow approaches a steady Couette flow with a linear spatially distributed velocity profile for slow oscillation.
	Consequently, the spin accumulation has also spatially uniform distribution, and the spin current cannot be generated by SVC.
	The oscillation frequency $\omega$ should be configured appropriately so that the boundary layer thickness is sufficiently shorter than the distance between the two plates to produce a spin current from an unsteady flow with the present system.

	\subsection{Transient flow}
\begin{figure}
	    \centering
	    \includegraphics[width=160mm]{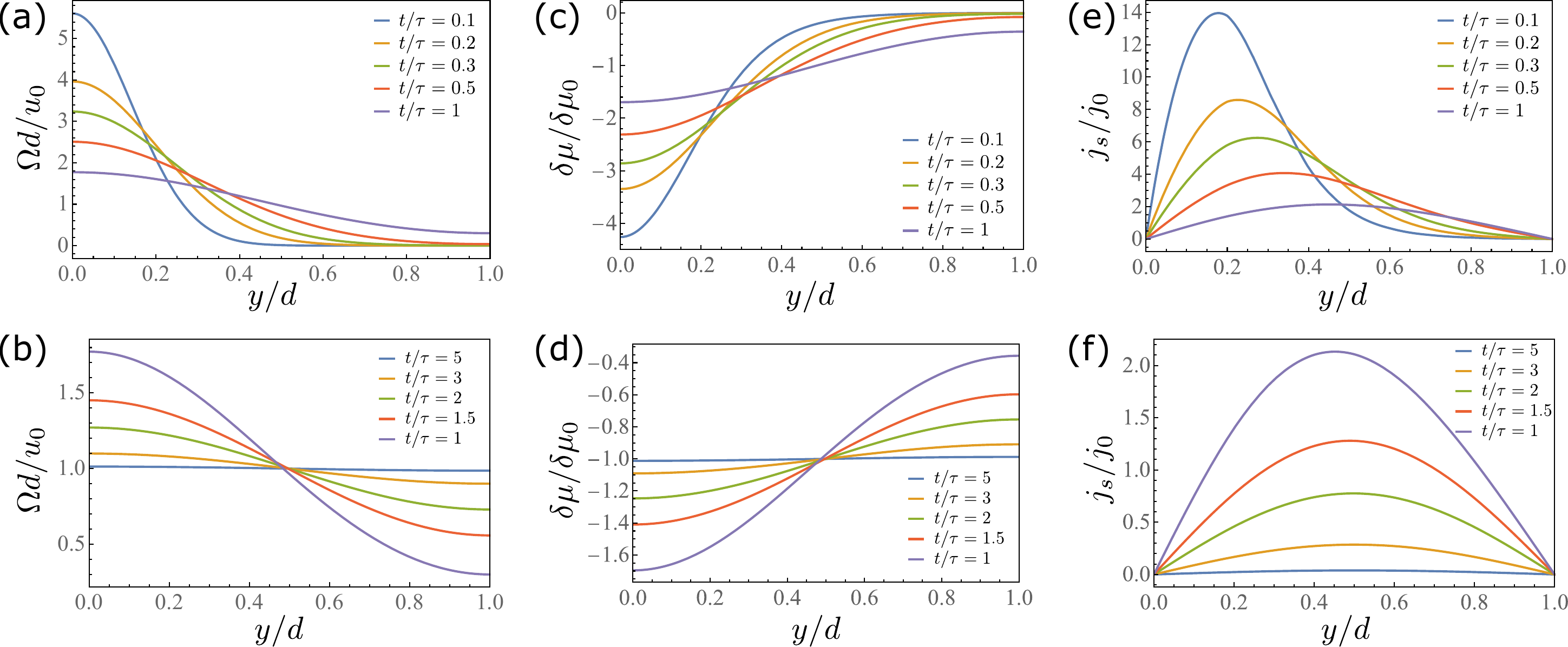}
	    \caption{
	    (Color online)
	    Spatial distribution of vorticity, spin accumulation, and spin current generated by the transient flow.
	    (a-b) The normalized vorticity, (c-d) the spin accumulation normalized by $\delta \mu_0=\hbar \zeta u_0 /d$, and (e-f) the spin current normalized by $j_0 = \sigma_c \hbar \zeta u_0/ed^2$. 
	    Colored lines represent the elapsed time since the initiation of the bottom plate movement, where $\tau$ is the decay time of the transient flow.
	    The vorticity and spin accumulation transform in time to spatially uniform distributions, and the transient spin current is generated.}
	    \label{fig:tran}
	\end{figure}

	Here, we focus on a transient flow wherein stationary fluid bounded by two infinite parallel plates at $y=0$ and $y=d$ is suddenly given movement at a certain moment.
	The bottom plate is supposed to be brought suddenly to a velocity $u_0$ in its own plane and subsequently maintained at this velocity while holding the upper plate stationary.
    With the boundary conditions of the velocity distribution being $u(0,t)=u_0$ and $u(d,t)=0$ for $t\geq 0$ and the initial condition being $u(y,0)=0$ for $0\leq y\leq d$, the solution of the Navier--Stokes equation is given by
	\begin{align}
	    u(y,t) = u_0 \left( 1-\frac{y}{d} \right) - \frac{2u_0}{\pi} \sum_{n=1}^{\infty} \frac{1}{n} \text{exp} \left( -n^2\pi^2 \frac{\nu t}{d^2} \right) \sin \frac{n\pi y}{d},
	    \label{eq:transient}
	\end{align}
	and this transient flow is known as the Rayleigh flow.
	Here, the first term of $u(y,t)$ corresponds to the steady Couette flow generated between the two rigid plates in relative steady motion in its own plane, and the second term of $u(y,t)$ corresponds to the transient part of the velocity profile of transient flow, which significantly decreases with time. 
	Therefore, the velocity distribution of the transient flow asymptotically approaches that appropriate for the steady Couette flow.
	The rapidity where the series in Eq.~(\ref{eq:transient}) tend to zero increases with $n$, and the first term with $n=1$ remains for the longest time.
	The asymptotic departure from the steady state decays approximately exponentially with the decay time $\tau = d^2/\pi^2 \nu$.
	The vorticity associated with the transient flow is given by
	\begin{align}
	    \Omega(y,t) = \frac{u_0}{d} + \frac{2u_0}{d} \sum_{n=1}^{\infty} \text{exp}\left( -n^2\pi^2 \frac{\nu t}{d^2}\right) \cos \frac{n\pi y}{d}.
	    \label{eq:omega_transient}
	\end{align}
	The time variation of the normalized vorticity is plotted in Fig.~\ref{fig:tran} (a) and (b).
	As expressed in Eq.~(\ref{eq:omega_transient}), the spatial distribution of vorticity becomes uniform as the transient flow asymptotically approaches a steady Couette flow.
	This flow is also readily apparent in the spatial distribution of the normalized vorticity with each time in Fig.~\ref{fig:tran} (a) and (b).
	
	We then substitute the vorticity associated with the transient flow into the spin-diffusion equation Eq.~(\ref{eq:spin_diff}), to calculate the spin current driven by the transient flow via SVC.
	The decay time of the transient state is assumed to be much longer than the spin relaxation time and that time derivative terms are negligible.
	We solve the steady spin-diffusion equation in Eq.~(\ref{eq:spin_diff_osc}) with the boundary conditions that the spin current as the two plates vanish, that is, $\partial_y \delta \mu_s(y=0) = \partial_y \delta \mu_s(y=d) = 0$.
	The normalized spin accumulation with each time is plotted in Fig.~\ref{fig:tran} (c) and (d), and its spatial variation evolves in time to spatially uniform distributions.
	Therefore, the spin current generated by the transient flow, which is the diffusion of the induced spin accumulation, flows transiently as shown in Fig.~\ref{fig:tran} (e) and (f).

	\subsection{Inverse spin Hall voltage}
	Let us estimate the inverse spin Hall voltage due to the spin current generated by an oscillatory flow and a transient flow, which is known as SHDG.
	The spin current is converted into electric voltage because of the spin--orbit interaction in liquid metal.
	The inverse spin Hall voltage, parallel to the flow velocity (the $x$-direction) between length $L_x$, is given by
	\begin{align}
	    V_{\text{ISHE}} = \theta_{\text{SH}} L_x L_z \int_0^d j_s(y) dy
	    = \theta_{\text{SH}} L_x L_z \Bigl[ \delta \mu_s(0) -\delta \mu_s(d) \Bigr]
	    ,
	\end{align}
	where $L_z$ is the length of the plates in the $z$-direction and  $\theta _{\text{SH}}$ is the spin Hall angle of liquid metal.
	
	First, we focus on an oscillatory flow.
	Considering the two parallel plates of length $L_z=2\,$mm in the $z$-direction separated by $d=100\, \mu$m, the bottom plate oscillates harmonically with the frequency $\omega=100\,$kHz and velocity $u_0=0.1\,$m/s.
	Using the charge conductivity for mercury is $\sigma_c=1.01 \times 10^6\,(\Omega \text m)^{-1}$ and the renormalization factor is $\zeta \sim 10^2$, the magnitude of the spin current generated by the oscillatory flow is obtained as $j_{s0}\simeq 0.67\,$A/m${^2}$.
	By use of the kinetic viscosity $\nu = 1.2 \times 10^{-7}\,$m${}^2$s${}^{-1}$, the spin relaxation length $\lambda \sim 10\,$nm, the spin Hall angle $\theta _{\text{SH}} \sim 0.01$, 
	the inverse spin Hall voltage between length $L_x=100\,$mm in the $x$-direction is estimated as $V_{\text{ISHE}}^{\text{osc}}\sim 9\,$nV, suggesting that the oscillatory flow induce the observable voltage.
	
	Second, we consider the same two parallel plates as in the oscillatory flow case, with the bottom plate moving suddenly move at a constant velocity $u_0=0.1\,$m/s.
	The decay time of the transient flow is calculated as $\tau \simeq 8.45\times 10^{-3}\,$s.
	The inverse spin Hall voltage induced by the transient flow at $t=1.0 \times 10^{-4\,}$s is estimated to be $V_{\text{ISHE}}^{\text{tran}} \sim 2\,$nV.

	\section{Conclusion}
	This study theoretically investigated the spin current generation in unsteady flows of liquid metal.
	We considered the two types of liquid-metal flow between two infinite parallel plates: the first one is an oscillatory flow because of the in-plane harmonic motion of one plane, and the other one is a transient flow because of the sudden constant motion of one plane. 
	The vorticity fields associated with the two-type flows were calculated using the Navier--Stokes equation of an incompressible viscous fluid.
	The spin accumulation and spin current were computed by using the spin-diffusion equation with SVC because of the calculated vorticity fields.
	We estimated the inverse spin Hall voltage converted from the spin current in unsteady flows of liquid mercury.
	The results indicate that unsteady flows can induce an observable magnitude of voltage.
	Our results are expected to enable the implementation of fluid spin devices with unsteady flows and expand the fluid spintronics application region.

	\begin{acknowledgments}
	We would like to greatly acknowledge the continued support of Y. Nozaki.
	We also thank T. Horaguchi and H. Nakayama for daily discussions.
	The authors would like to thank MARUZEN-YUSHODO Co., Ltd. (https://kw.maruzen.co.jp/kousei-honyaku/ ) for the English language editing.
	This work was partially supported by JST CREST Grant No. JPMJCR19J4, Japan.
	This work was supprted by JSPS KAKENHI for Grant Nos. 20H01863, 20K03831, 21H04565, 21H01800, and 21K20356.
	MM was supported by the Priority Program of Chinese Academy of Sciences, Grant No. XDB28000000.
	\end{acknowledgments}
	
%\bibliography{shd.bib}

	\end{document}